\documentclass[a4,12pt]{article}
\usepackage[dvips]{graphicx}
\usepackage{amsmath,amssymb}
\usepackage{amssymb}
\usepackage{enumerate}
\usepackage[usenames]{color}

\renewcommand{\thefootnote}{\fnsymbol{footnote}}

\begin{document}
\title{
Non-Abelian Strings in High Density QCD:\\
Zero Modes and Interactions}
\maketitle

\begin{center}
\author{
Eiji Nakano$^a$\footnote{e-mail: {\tt enakano(at)ntu.edu.tw}}, 
Muneto Nitta$^b$\footnote{e-mail: {\tt nitta(at)phys-h.keio.ac.jp}}
 and 
Taeko Matsuura$^c$\footnote{e-mail: {\tt taeko(at)ect.it}}
}

\bigskip\bigskip\bigskip
$^a$ {\it Department of Physics and Center for
Theoretical Sciences, National Taiwan University, Taipei 10617, Taiwan}\\

$^b$ {\it Department of Physics, Keio University, Hiyoshi, Yokohama,
Kanagawa 223-8521, Japan}\\

$^c$ {\it ECT*, Villa Tambosi, 
        strada delle Tabarelle, 286, I 38050 Villazzano (TN), Italy}\\

\abstract
{ The most fundamental strings 
in high density 
color superconductivity are the non-Abelian semi-superfluid strings 
which have color gauge flux tube but behave as 
superfluid vortices in the energetic point of view.
We show that in addition to 
the usual translational zero modes, 
these vortices have normalizable orientational zero modes 
in the internal space, 
associated with 
the color-flavor locking symmetry 
broken in the presence of the strings.
The interaction among two parallel non-Abelian 
semi-superfluid strings 
is derived for general relative 
orientational zero modes 
to show the universal repulsion. 
This implies that the previously known superfluid vortices, 
formed by spontaneously broken $U(1)_{\rm B}$, are unstable to decay.
Moreover, our result proves the stability of color superconductors 
in the presence of external color gauge fields.
}
\end{center}
\newpage

\setcounter{footnote}{0}
\renewcommand{\thefootnote}{\arabic{footnote}}

%
\section{Introduction}
%
\indent

%
%

%
%

Quark matter at high density is considered to exhibit 
color superconductivity \cite{colorsc}. 
There are lots of phases proposed in the 
color superconductivity and still there seems no agreement 
on which is the true ground state \cite{various}. 
However, when the quark chemical potential 
becomes much larger than the strange quark mass $\mu \gg m_s$, 
the color-flavor locking (CFL) phase is expected to 
be realized \cite{Alford:1998mk}. 
In the CFL phase, the quark-quark pairing breaks the symmetry
of QCD such as $G=U(1)_{\rm B} \times SU(N)_{\rm C} \times SU(N)_{\rm F}$ 
$\rightarrow$ $H=SU(N)_{\rm C+F} \times {\bf Z}_N$ with $N=3$,
where $U(1)_{\rm B}$ is a baryon number symmetry and 
$SU(N)_{\rm C}$, $SU(N)_{\rm F}$ and $SU(N)_{\rm C+F}$ are 
color, flavor and the color-flavor locked symmetries, respectively. 
The breaking of $U(1)_{\rm B}$ produces the superfluid vortices 
\cite{Forbes:2001gj,Iida:2002ev} which may play a role in
the neutron star physics. 
It has been, however, shown in Ref.~\cite{Balachandran:2005ev} 
that they are not the fundamental strings in
the color superconductivity.
The most fundamental strings are then the semi-superfluid strings
which are {\it non-Abelian} strings with color gauge flux tube.
A superfluid vortex in three flavor QCD 
can be topologically (and group-theoretically) 
decomposed into three non-Abelian strings.
It remains as a significant open problem which is really 
energetically favored,
three separated non-Abelian vortices or
one superfluid vortex as a bound state of them. 

%
%

In general,
non-Abelian strings are strings 
which arise for symmetry breaking $G \to H$ 
in which the unbroken subgroup $H$ is non-Abelian.
This kind of strings themselves attracts lots of attention these 
several years. 
Recently the non-Abelian local strings have been found 
in superstring theory \cite{Hanany:2003hp} and in supersymmetric QCD 
\cite{Auzzi:2003fs}. 
For instance, they have been used to show 
color confinement and non-Abelian duality 
in supersymmetric QCD \cite{Eto:2006dx}.
Since these strings are BPS, namely at critical coupling, 
there exists no static force between them and therefore
the moduli space is admitted.
The most generic solutions
and their moduli space have been obtained \cite{Eto:2005yh}
by introducing the method of the moduli matrix 
\cite{Isozumi:2004vg,Eto:2006pg}. 
Dynamics of strings such as 
reconnection (intercommutation) of two strings
has been studied using the moduli space approximation \cite{Eto:2006db}. 
Non-Abelian semi-local strings have also been
extensively studied \cite{Eto:2006pg,Shifman:2006kd}. 
As for non-Abelian global strings, 
there has been not so much work, but they may be generated
during the chiral phase transition in high temperature QCD
\cite{Balachandran:2002je, NS, NNM}.

%
%

Different from Abelian strings, 
the distinct character of non-Abelian strings is 
their internal degrees of freedom which are called $orientation$.
In the case of generalized QCD where 
the number of flavor equals to the number of flavor,
the presence of a string breaks the symmetry $H$ further as 
$SU(N)_{\rm C+F} \to SU(N-1)_{\rm C+F} \times U(1)_{\rm C+F}$. 
Consequently the zero modes corresponding to 
\begin{eqnarray}
 {SU(N)_{\rm C+F} \over SU(N-1)_{\rm C+F} \times U(1)_{\rm C+F}} 
 \simeq {\bf C}P^{N-1}  \label{eq:orientation}
\end{eqnarray}
appear along the string. 
Then we have a continuously
infinite number of strings 
with the same tension, 
which are parameterized by 
the $orientation$, namely a point in ${\bf C}P^{N-1}$.
In the case of local (global) non-Abelian strings, 
these orientational zero modes are (non-)normalizable 
because the transformation fixes \cite{Auzzi:2003fs} 
(changes \cite{NS,NNM}) the boundary condition 
of the strings at infinity. 
In the case of semi-superfluid strings, 
it is, however, a non-trivial problem whether 
orientations are normalizable or not, because 
$U(1)_{\rm B}$ symmetry is a global symmetry, 
unlike the case of local strings \cite{Auzzi:2003fs} 
where $U(1)_{\rm B}$ is gauged.

In this Letter, we first show that 
orientational zero modes are in fact normalizable. 
We then calculate
the force among 
two non-Abelian semi-superfluid strings 
with general orientations. 
We find that the static force is always repulsive and is 
$1/N$ of that between superfluid vortices. 
It does not depend on the orientations of the strings,  
which is somewhat surprising because 
the static force between two global non-Abelian strings 
{\it does} depend on the orientations \cite{NNM}.
Our result implies that a superfluid vortex 
of $U(1)_{\rm B}$ breaking found in Ref.~\cite{Forbes:2001gj,Iida:2002ev}
is actually unstable to decay into 
$N$ (three) non-Abelian strings by repulsive force between them. 

In the case of the usual superconductors, 
the interaction between two strings were important; 
the repulsion (attraction) between strings 
in type II (I) superconductors 
implies their (in)stability in the presence of 
an external magnetic field.
The universal repulsion found in this Letter 
ensures the stability of color superconductors
in the presence of external color gauge fields,  
regardless of whether they are of type I or II.

%
%

The construction of this Letter is the following.
In section 2, we review the Ginzburg-Landau 
Lagrangian which has 
generalized QCD symmetry $G$ and the construction of 
single non-Abelian string.
In section 3, the non-Abelian string solution with 
general orientation is constructed.  
The interaction among the static two non-Abelian 
strings with general relative orientations 
is derived in section 4. 
We end in section 5 with conclusion and discussion.

\section{Non-Abelian semi-superfluid strings}
%
Let us start from constructing the general form of the Ginzburg-Landau
Lagrangian \cite{GL} on the basis of the generalized QCD symmetry:
\begin{eqnarray}
G=SU(N)_{\rm C} \times SU(N)_{\rm F} \times U(1)_{\rm B},
\end{eqnarray}
where we set the number of flavor equals to the number of flavor 
and $SU(N)_{\rm C}$ is the gauge symmetry. 
For this purpose, we first introduce 
an $N$ by $N$ matrix field $\Phi_{\alpha i}$ 
($\alpha, i=1, \cdots, N$) 
where $\alpha(i)$ denote the color(flavor) indices. 
$\Phi$ belongs to $[N, N]$ representation of 
$SU(N)_{\rm C} \times SU(N)_{\rm F}$. 
The symmetry $G$ transforms the matrix field as 
\begin{eqnarray}
\Phi \to e^{i\alpha} U_{\rm C} \Phi U_{\rm F}^{t},
\end{eqnarray}
where $U_{\rm C}$ and $U_{\rm F}$ are independent $SU(N)$ matrices 
and $e^{i\alpha}$ is a global $U(1)_{\rm B}$ rotation 
associated with the baryon number conservation.

In the case of color superconductivity, 
$\Phi$ field corresponds to the pairing gap; 
the spin-zero pairing of the positive energy quarks 
in antisymmetric combinations of colors and flavors \cite{Pisarski:1999tv}. 
In this case, we may write 
$\Phi$ as 
\begin{eqnarray}
 \Phi_{\alpha i} \sim  \epsilon_{\alpha\beta\gamma}
\epsilon_{ijk} \langle \psi^{T\beta}_{j} C \gamma_5 \psi^{\gamma}_{k} \rangle,
\end{eqnarray}
where $\psi$ is the quark field and we assumed that the ground state 
is the positive parity state which would be determined by the instanton effect. 
Then the most general 3-d Ginzburg-Landau Lagrangian up to 
${\cal O}(\Phi^4)$ is:
\begin{eqnarray}\label{GL}
{\cal L} =
 {\rm{tr}}({D}\Phi)^{\dag}({D}\Phi)
-m^2{\rm{tr}} (\Phi^{\dag}\Phi)
-\lambda_1 ( {\rm{tr}}\Phi^{\dag}\Phi)^2
-\lambda_2 {\rm{tr}}\left[(\Phi^{\dag}\Phi)^2\right] 
-\frac{1}{4}F^a_{ij}F^{a   ij}, 
\end{eqnarray}
where $D\equiv \partial -i g_s A^\alpha T_\alpha$ is 
the covariant derivative for the color symmetry, 
and $T_a$ $(a=1,2, \cdots, N^2-1)$ are the generators of $SU(N)_{\rm C}$ 
in the fundamental representation which we normalize as 
${\rm Tr}\{T_aT_b\}=\delta_{ab}$.
Here $g_s$ is the gauge coupling and 
$ F_{ij}^a = \partial_i A_j^a - \partial_j A_i^a
      + g_s f_{abc}A_i^b A_j^c$ is the field strength.\footnote{
Throughout this Letter we do not consider the electro-magnetic (EM) 
symmetry for simplicity.
The EM group $U(1)_{\rm EM}$ is a subgroup of the flavor symmetry 
$SU(N)_{\rm F}$ 
and explicitly breaks $SU(N)_{\rm F}$.
The zero modes found in this Letter are thus 
massive through the EM interaction.
}

Stability condition of vacua 
enforces $\lambda_1 + \lambda_2 /N >0$. 
When $m^2 <0$ and $\lambda_2 >0$, 
the vacuum expectation value takes 
\begin{eqnarray}
\langle \Phi \rangle = v {\bf 1} \equiv \Phi_0, 
~ v = \sqrt{-m^2/2(N \lambda_1 + \lambda_2)}, 
\end{eqnarray}
and the symmetry $G$ is broken to $H=SU(N)_{\rm C+F} \times {\bf Z}_N$. 
This condensation is called the color-flavor locking (CFL) 
in the case of the color superconductivity.
The action of $H$ on $\langle \Phi \rangle$ is 
$\langle \Phi \rangle \to e^{i\alpha} U_{\rm C}
\langle \Phi \rangle U_{\rm F}^{t}$ with 
$(e^{i\alpha}, U_{\rm C}, U_{\rm F}^{t})
=(\omega ,\omega^{-1} U, U^{\dag}):
\omega \in {\bf Z}_N, U \in SU(N)$. 
The coset space $G/H \simeq U(N)$ has 
the non-trivial first homotopy group $\pi_1[U(N)]\simeq {\bf Z}$, 
which develops non-Abelian as well as Abelian strings. 
In the vacua, only $U(1)_{\rm B}$ Nambu-Goldstone boson 
remains massless with the rests Higgssed. From now on, 
we will take $v=1$ for simplicity.

Here we concentrate on the most fundamental string 
out of which all the other string configurations are made, 
and it actually has the lowest energy configuration with a non-trivial loop. 
Since the common element 
${\bf Z}_N$ of $SU(N)_{\rm C, F}$ and $U(1)_{\rm B}$ 
provides the warp points to make the non-trivial loops therein, 
the fundamental string (non-Abelian string) is generated by 
{\it both} $SU(N)_{\rm C, F}$ and $U(1)_{\rm B}$ generators. 

To make this statement clear, we consider the cylindrically 
symmetric string configuration along the $z$-axis. 
In the polar coordinates ($\rho, \theta$) in the $x$-$y$ plane, 
an isolated fundamental string has the form 
\begin{eqnarray}\label{reference1}
\Phi(\theta, \rho)&=&  \exp \left(i \frac{\theta}{N}\right) 
\exp \left(-iT_{N^2-1} \frac{\sqrt{N(N-1)}}{N}\theta \right) 
{\rm diag}\left(f(\rho), g(\rho), \cdots , g(\rho) \right)
\nonumber\\
&=& {\rm diag}(e^{i\theta} f, g, \cdots , g)
\end{eqnarray}
We take the basis so that the ($N^2-1$)-th generator is 
$T_{N^2-1}= \frac{1}{\sqrt{N(N-1)}}{\rm diag}(1- N , 1,\cdots ,1)$. 
$f$ and $g$ are functions of $\rho$, 
and $f(0)=0$ and $f(\infty)=g(\infty)=1$ at boundaries. 
At sufficiently large distance from the core, $\rho \gg \lambda$ 
where $\lambda \equiv m^{-1}$ is the coherence length, 
the profile of the string is well described by 
\begin{eqnarray}\label{reference}
\Phi(\theta, \rho)
 \simeq  {\rm diag}(e^{i\theta}, 1, \cdots, 1) 
\quad \mbox{for} \quad \rho \gg \lambda. 
\end{eqnarray}
From Eq.~(\ref{reference1}), 
one can see that the non-trivial loop
must be made from both generators of $SU(N)$ and $U(1)_{\rm B}$.
Although from topological reason we could use either $SU(N)_{\rm C}$ or 
$SU(N)_{\rm F}$ or both of them for $SU(N)$, the energy consideration forces 
us to use the gauge symmetry $SU(N)_{\rm C}$ to minimize the energy of the 
covariant derivative term. 
Therefore, we use both global $U(1)_{\rm B}$ and local $SU(N)_{\rm C}$ 
symmetries to construct the string.
The name {\it semi-superfluid} originates this fact.

The gauge fields associated with the string is determined 
by minimizing the kinetic energy as much as they can. 
Note that, 
different from local strings, the covariant derivative term 
does not vanish at infinity but it is finite, which makes 
the energy of the string logarithmically divergent 
as the system size is infinite,
just like superfluid vortices. 
Given Eq.~(\ref{reference}), the form of the gauge field can be deduced as:
\begin{eqnarray}\label{reference2}
  A_\theta^{N^2-1} =  \frac{\xi h(\rho)}{g_s \rho}, 
\end{eqnarray} 
and the other components of the gauge field vanish. 
Here the gauge field only has non-zero 
component for $\theta$ direction, which we have denoted as $A_\theta$.  
The function $h(\rho)$ satisfies $h(\infty)=1$ at boundary. 
A constant $\xi$ should be determined so as to minimize 
the kinetic energy density $F_{\rm kin}$ at infinity: 
\begin{eqnarray}\label{xi1}
F_{\rm kin}={\rm tr}\left|D \Phi \right|^2={\rm tr} 
\left| \left( \frac{1}{\rho}
\frac{\partial}{\partial \theta} -i g_s A_\theta^{N^2-1} T_{N^2-1} \right) \Phi \right|^2
=\frac{1}{\rho^2} 
\left[ \left( \xi + \sqrt{\frac{N-1}{N}} \right)^2 +\frac{1}{N} \right], 
\end{eqnarray}
and its minimization is achieved at 
\begin{eqnarray}\label{xi2}
\xi=-\sqrt{\frac{N-1}{N}}. 
\end{eqnarray}
The kinetic energy then becomes $F_{\rm kin}=\frac{1}{N\rho^2}$,
which is $1/N$ of that 
of the global $U(1)_{\rm B}$ string \cite{Forbes:2001gj}.
Thus, at large distance over the penetration depth 
$\lambda_v \equiv m_g^{-1} 
\sim g_s^{-1} v^{-1}$, 
the gauge field configuration becomes 
\begin{eqnarray}\label{gauge1}
ig_s A_\theta^{N^2-1} T_{N^2-1}
\simeq -\frac{i}{N \rho}
\left(\begin{array}{cc}
1-N& 0 \\ 
0 & {\bf 1}_{N-1} 
 \end{array} \right)
\quad \mbox{for} \quad \rho \gg \lambda_v. 
\end{eqnarray}

The numerical solution of $f$, $g$, and $h$ 
for the semi-superfluid string 
in the color-flavor locked phase is found in Ref.~\cite{Balachandran:2005ev}, 
where the color and electro-magnetic fields are mixed 
and only one of their linear combinations is relevant for flux.

\section{Internal space and color gauge transformation}
%
Before going ahead, it is instructive to clarify 
the internal space of the string. 
The presence of the string (\ref{reference1}) breaks 
the symmetry $H$ further to 
$SU(N)_{\rm C+F} \to SU(N-1)_{\rm C+F} \times U(1)_{\rm C+F}$. 
The internal space corresponds to the coset space  
(\ref{eq:orientation}).
Zero modes parameterizing the space (\ref{eq:orientation}) 
appear along the string, i.e, 
Eq.~(\ref{reference}) denotes just one particular string 
of a continuously infinite number of strings with 
the same string tension (flux energy) 
which are parameterized by the {\it orientation} in the ${\bf C}P^{N-1}$. 
Unlike the case of global strings \cite{NS,NNM} 
these zero modes are normalizable as seen below.

Here we consider a fundamental string with general orientation 
in the internal space. 
We first take the fundamental string Eq.~(\ref{reference1}) 
as a reference string $\phi_0$, 
\begin{eqnarray}\label{phi1}
 \phi_0  = {\rm diag}(e^{i\theta}f,g, \cdots , g, g).
\end{eqnarray} 
Then the string  $\phi$ with general orientation 
in ${\bf C}P^{N-1}$ relative to the reference string should be obtained 
by $SU(N)_{\rm C+F}$ transformation to $\phi_0$. 
However, there are some redundancies in this transformation, i.e., 
only an $SU(2)_{\rm C+F} \left( \subset SU(N)_{\rm C+F} \right)$ 
rotation is enough to be considered for relative orientation to $\phi_0$  
without loss of generality.
This corresponds to a ${\bf C}P^1$ submanifold in the whole ${\bf C}P^{N-1}$. 
Furthermore, 
since any regular color-gauge transformation does not 
change the physical situation, 
we omit $SU(2)_{\rm C}$ rotation to $\phi_0$ for the moment. 
Thus we transform $\phi_0$ by an element $U_{\rm F}$ of flavor $SU(2)_{\rm F}$:
\begin{eqnarray}
 \phi 
 = \phi_0 U_{\rm F}^{t}
 =\left(\begin{array}{cc}
   \left(\begin{array}{cc}
     e^{i\theta} f & 0 \\
                 0 & g \\  
   \end{array} \right)
  u_{\rm F}^{-1} & 0 \\
               0 & g {\bf 1}_{N-2} 
 \end{array} \right)
= 
 \left(\begin{array}{cc}
  \left(\begin{array}{cc}
   e^{i\theta} a f & e^{i\theta} b f \\
            -b^* g & a^* g \\  
  \end{array} \right) & 0  \\
                    0 & g {\bf 1}_{N-2}  \\
 \end{array} \right),
  \label{phigene}
\end{eqnarray}
where 
$u_{\rm F} \equiv 
\left(\begin{array}{cc}
a^* & -b \\
b^* & a \\  
\end{array} \right)$
(with $|a|^2 + |b|^2 = 1$) is an element of $SU(2)_{\rm F}$. 
This is a general expression for the fundamental string.  
However, as will be shown below, 
the flavor-rotated string configuration (\ref{phigene}) can be transformed 
either back to the original form $\phi \simeq {\rm diag}(e^{i\theta},1, \cdots, 1)$, 
or to the form $\phi \simeq {\rm diag}(1, e^{i\theta},1, \cdots, 1)$ 
with fully opposite orientation, 
at a region away from the centre of strings, 
by use of a {\it twisted} color-gauge transformation. 

Any color gauge transformation keeps 
the physical situation unchanged if they are regular. 
Here we implement a {\it twisted} color transformation 
of $SU(2)_{\rm C}$, given by
\begin{equation}\label{twist1}
u_{\rm C}(\theta, \rho)=
\left(\begin{array}{cc}
a^* & -b e^{i \theta F(\rho) }  \\
b^* e^{-i \theta F(\rho)}  & a 
\end{array} \right) 
\end{equation}
with $F(\rho)$ being an arbitrary regular function 
with boundary conditions  $F(0)=0$ and $F(\infty)=1$. 
The former condition has been imposed 
to make the transformation regular at the center of string. 
This is possible because of 
$\pi_1 [SU(2)_{\rm C}] = 0$.
The upper left 
$2 \times 2$ minor matrix of 
$\phi$ in (\ref{phigene}) is transformed to 
\begin{eqnarray}\label{phi2twist1}
u_{\rm C}(\rho, \theta)
\left(\begin{array}{cc}
e^{i\theta} a f & e^{i\theta} b f \\
-b^* g & a^* g \\  
\end{array} \right)
&=&
\left(\begin{array}{cc}
  |a|^2 f e^{i \theta}+|b|^2 g e^{i\theta F} 
& a^* b \left[-e^{i\theta F}+f e^{i\theta} \right] \\
a b^* \left[-1+f e^{i(1-F)\theta} \right] 
&|a|^2 g+|b|^2 f e^{i(1-F)\theta }   
 \end{array} \right) \nonumber \\
&\simeq&  
\left(\begin{array}{cc}
 e^{i \theta} & 0 \\
 0 & 1 
 \end{array} \right) 
\quad \mbox{for} \quad \rho \gg \lambda. 
\end{eqnarray}
This result means $\phi \simeq \phi_0$ 
for $\rho \gg \lambda$. 
Also, the fully opposite orientation can be obtained 
by another color-gauge transformation, 
\begin{equation}\label{twist2}
\left( 
\begin{array}{cc}
e^{-i \theta F(\rho)}  & 0 \\
0 & e^{i \theta F(\rho)} 
 \end{array} 
\right)
u_{\rm C}
\left( 
\begin{array}{cc}
f e^{i \theta}  & 0 \\
0 & g 
 \end{array} 
\right)u_F^{-1} 
\simeq 
\left( 
\begin{array}{cc}
1  & 0 \\
0 &  e^{i \theta}
 \end{array} 
\right) 
\quad \mbox{for} \quad {\rho \gg \lambda}. 
\end{equation}
Note that one cannot change the topological number 
by use of this kind of regular gauge transformations. 
We thus have seen that spatial infinity of the string configurations 
is the same and does not depend on the orientational zero modes.
This implies that the orientational zero modes are {\it normalizable}, 
unlike the case of global strings \cite{NS,NNM}.

In conclusion, 
all semi-superfluid strings with general orientation 
are equivalent to each other away from the core. 
In other words, 
strings rotated by the flavor $SU(2)_{\rm F}$ revert to 
the first fundamental string given by Eq.~(\ref{reference}) 
via the color gauge transformation at much longer distances than 
the coherence length. 
This fact makes the problem of the static long range force 
between two strings significantly simple.

\section{Interaction between two strings}
\begin{figure}
\begin{center}
\includegraphics[height=3.0cm, width=6cm]{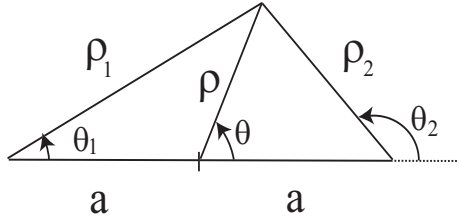}
\end{center}
\caption{\label{fig1} 
Configuration of two semi-superfluid strings with interval $2a$ 
in the polar coordinates $(\rho, \theta)$.
$\rho_{1,2}$ is distance from string $\phi_{1,2}$, 
and $\theta_{1,2}$ is angle around it. }
\end{figure}

%
Here we consider the interaction between arbitrary two strings 
named $\phi_1$ and $\phi_2$.
These strings are placed at $(\rho, \theta)=(a,\pi)$ and $(a,0)$ 
in parallel along the $z$-axis, see Fig.~\ref{fig1}. 
We eventually go for 
the expression of a long range static force between two strings, 
which is valid if string are sufficiently separated. 
The situation we consider is sorted out as follows:  
\begin{itemize}
\item The interval between strings is much larger than both 
the coherence length 
and the penetration depth: 
$a \gg \lambda (= m^{-1}), \lambda_v (= g_s^{-1} v^{-1})$. 

\item The first string $\phi_{1}$ 
is approximated everywhere by the asymptotic profile (\ref{reference}): 
$\phi_{1}={\rm diag}\left( e^{i\theta_{1}}, 1, \cdots, 1 \right)$. 
%
The second string $\phi_{2}$ has the profile (\ref{phigene}) 
with general orientation relative to $\phi_1$. 
At large distance of our interest, however, 
it is equivalent to the reference string configuration (\ref{reference}): 
$\phi_{2} \simeq {\rm diag}\left( e^{i\theta_{2}}, 1, \cdots, 1 \right)$. 
$\phi_{1,2}$ becomes an anti-string by changing the signs of $\theta_{1,2}$.

\item The total profile of the two string system is given 
by the Abrikosov ansatz: 
$\Phi_{\rm tot}=\phi_1\phi_2$ 
and 
$A^{\theta}_{\rm tot}=A^{\theta}_1+A^{\theta}_2$.
The first ansatz does not depend on the ordering of the matrices 
because the second string transforms to diagonal at large distance 
as shown in Sec.~3.\footnote{
One can easily show that the {\it twisted} color transformation also works 
to make a product 
(\ref{phi1}) $\times$ (\ref{phigene}) 
as $\phi_1 \times \phi_2$ diagonalized at large distance.
}
$A^{\theta}_{1,2}$ is the gauge field configuration (\ref{gauge1}) 
accompanied with the single string system of $\phi_{1,2}$. 
For an anti-string, $A^{\theta}_{1,2}$ changes the signs.

\end{itemize}

Now we are ready to evaluate the interaction between two 
parallel non-Abelian strings with general orientations 
in the internal space.
In order to obtain the static force between them, 
we first calculate the interaction energy density of the two string system, 
which is obtained by subtracting two individual string energies 
from the total configuration energy. 
According to the above situation, 
the interaction energy density is given as 
\begin{eqnarray}\label{F}
F(\rho,\theta,a)
&\simeq&  {\rm tr}
\left( |D \Phi_{\rm tot}|^2 -|D \phi_1|^2
-|D \phi_2|^2  \right) \nonumber \\
&=& \pm
\frac{2}{N}
\left[ 
\frac{-a^2 + \rho^2}
{a^4 + \rho^4 -2 a^2 \rho^2 \cos (2 \theta)}
\right], 
\end{eqnarray}
where we have used the fact that $V(\Phi_{\rm tot})=V(\phi_1)=V(\phi_2)=0$ 
and $F_{ij}^aF^{aij}=0$ at large distance \cite{Perivolaropoulos:1991du}. 
Here and below, the upper(lower) sign indicates the quantity 
for string-string(string-anti-string) configuration.

The tension, the energy of the string per unit length,  
is obtained by integrating the energy density
over the $x$-$y$ plane, 
\begin{eqnarray}\label{E}
E(a,L)= \pm
\int _{0}^{L} d\rho 
\int _{0}^{2 \pi} d\theta  \rho F(\rho,\theta,a)
= \pm \frac{2 \pi}{N} 
\left[ -\ln 4 -2 \ln a  + \ln \left(a^2 + L^2 \right)  \right],
\label{eq:int-energy}
\end{eqnarray}
where the IR cutoff $L$ is introduced to make the integral finite. 
The force between the two strings are then obtained by 
differentiating $E$ by the interval:
\begin{eqnarray}\label{f}
 f(a,L)
 = {\mp} 
 \frac{\partial E}{2 \partial a}
 =  \pm \frac{2 \pi}{N} 
\left(\frac{1}{a} - \frac{a}{a^2 + L^2}  \right) 
\simeq  \pm \frac{ 2 \pi }{N a},
\end{eqnarray}
where the last expression is for $a \ll L \to \infty$.
We can see that the force is {\it repulsive(attractive)} 
for string-string(anti-string-string) configuration. 
The overall factors $1/N$ in Eqs.~(\ref{F})--(\ref{f}) are attributed 
to the fact 
that the tension of the fundamental non-Abelian string 
is reduced by $1/N$ compared to the usual Abelian string, 
then leading to $1/N$ erosion in magnitude of the force.

Note that our result does not depend on whether the superconductivity 
is of type I or II. 
This has an important meaning in the case of color superconductivity
since although the perturbation theory indicates 
the color superconductivity is of type I for whole density regime
\cite{Giannakis:2003am},
the most fundamental strings, semi-superfluid strings, 
can be stable at any density regime where CFL realizes.
This result also implies that 
the global $U(1)_{\rm B}$ superfluid strings 
$\Phi \simeq {\rm diag} (e^{i \theta},\cdots,e^{i \theta})$ 
found in Ref.~\cite{Forbes:2001gj,Iida:2002ev} 
as well as the $M_2$ strings 
$\Phi \simeq {\rm diag} (1,e^{-i \theta},e^{-i \theta})
\simeq (e^{2i \theta},1,1)$ 
suggested in Ref.~\cite{Balachandran:2005ev} 
are unstable to decay into 
$N$ or 2 semi-superfluid strings, respectively. 
It contrasts to the case of global non-Abelian strings \cite{NNM}, 
where the $U(1)$ Abelian string is marginally unstable, 
$i.e.$, 
no force exists between two strings with opposite orientations.

From all the above arguments 
we conclude that there exists the long range force 
between two semi-superfluid strings, 
which is independent of the orientations in the internal space,
and the sign of the force is determined 
by the difference in their topological charges.

\section{Discussion and Outlook}
%
\indent
In this Letter, we have considered the interaction between two non-Abelian 
semi-superfluid strings in the system which has generalized QCD symmetry, 
$SU(N)_{\rm C} \times SU(N)_{\rm F} \times U(1)_{\rm B}$, 
using the Abrikosov approximation.  
This approximation is justified for 
the case where the strings are far apart.
In case of a short interval, however, 
where the string cores overlaps: $a \sim \lambda, \lambda_v$, 
the present treatment might not be applicable 
to describe a fine structure of the force. 
In such a case the Abrikosov ansatz for $\Phi_{\rm tot}$ 
we have employed might be suspicious, 
since amplitude functions $f$ and $g$ veer away from the unity near the core, 
and arbitrary orientation in the internal space 
made by flavor $SU(2)_{\rm F}$ rotation 
generates off-diagonal parts in the string configuration. 
It means that two strings do not commute: 
$\left[ \phi_1, \phi_2 \right] \neq 0$. 
However our conclusion of the instability of $U(1)_{\rm B}$ strings 
(or the stability of color superconductors) 
is unchanged.
Even if the short range force is attractive 
a $U(1)_{\rm B}$ string will decay by  
long range repulsion through classical large fluctuations, 
thermal fluctuations  
or the quantum tunnelling effect. 

As applications of our results to the compact star physics, 
the universal repulsion implies that 
the lattice structure of many-string system may be obtained
during the rapid cooling of the protoneutron stars or  
in response to the external electro-magnetic field 
and/or the rotation. 
This might have impacts on observables 
such as the pulsar glitch phenomenon. 
It is, however, still an open question how the strings 
terminate at the interface between 
the color superconductor in the core and
the surrounding nuclear matter.

Another interesting issue is how the non-Abelian string 
releases or interacts with the Nambu-Goldstone bosons corresponding to
the global $U(1)_{\rm B}$ breaking as well 
as the leptons, quarks, mesons etc   
which exist in the neutron star. 
The former may be described using the two index antisymmetric 
tensor representation in which the Kalb-Ramond action appropriately 
describes the string \cite{Kalb:1974yc}. 
Interaction between strings and other topological solitons 
is also interesting to be explored. 
It has been shown in Ref.~\cite{Son:2000fh} 
that $U(1)_{\rm A}$ domain walls appear 
when the anomalous $U(1)_{\rm A}$ is spontaneously broken.
Fundamental quarks appear as Skyrmions (called qualitons) 
in the CFL phase \cite{Hong:1999dk}.
Interaction of non-Abelian strings and these objects 
remains an open problem.

When the density is decreased, the various kinds of phases
may appear due to the strange quark mass and the electric 
neutrality conditions; 2SC, dSC, uSC, gluonic phase, meson condensed phases,
gapless phases, FFLO etc. There might appear new topological 
objects in these phases 
\cite{Kaplan:2001hh,Gorbar:2005pi}.
In particular, 
it has been shown in Ref.~\cite{Kaplan:2001hh} that 
there appear $K$-strings, drum vorton and domain walls 
in one of meson condensed phases, the CFL + $K^0$ phase.
Our work should be applied to the interaction of these strings.


NOTE: 
The non-Abelian strings discussed in a recent paper 
\cite{Gorsky:2007kv} are completely different from our strings 
appearing in high density QCD. 
Their strings are local strings with gauged $U(1)_{\rm B}$ 
and are essentially the same with \cite{Hanany:2003hp,Auzzi:2003fs}.

\section*{Acknowledgements}
T.M. is very grateful to S. Digal for helpful comments 
and encouragements. 
The work of E.N. is supported by 
Center for Theoretical Sciences, National Taiwan University 
under grant NO.~NSC96-2811-M-002-024. 
E.N. would like to thank theoretical high energy group 
at Tokyo Institute of Technology for kind hospitality. 


\end{document}